\documentclass{svmult}

\usepackage{float}
\restylefloat{figure}

\usepackage{makeidx}         
\usepackage{graphicx}        
\usepackage{multicol}        
\usepackage{subfig}
\usepackage[bottom]{footmisc}

\usepackage{enumerate}
\usepackage[numbers,sort&compress]{natbib}

\usepackage{amsxtra}     
\usepackage{amssymb} 
\usepackage{amsfonts}

\makeindex             

\usepackage{caption}

\begin{document}

\title*{Modeling the desired direction in a force-based model for 
pedestrian dynamics}
\titlerunning{Modeling the desired direction in a force-based model} 
\author{Mohcine Chraibi\inst{1,2}\and
Martina Freialdenhoven\inst{1}\and
Andreas Schadschneider\inst{3}\and 
Armin Seyfried\inst{1,2}
}
\authorrunning{M. Chraibi, M. Freialdenhoven, A. Schadschneider and A. Seyfried}
\institute{J\"ulich Supercomputing Centre, Forschungszentrum J\"ulich.
52425 J\"ulich, Germany.\\
\texttt{m.chraibi@fz-juelich.de}\\
\texttt{MartinaFreialdenhoven@web.de}\\
\texttt{a.seyfried@fz-juelich.de}\\
\and Computer Simulation for Fire Safety and Pedestrian Traffic, 
Bergische Universit\"at Wuppertal, Pauluskirchstra{\ss}e 11, 
D-42285 Wuppertal, Germany. 
\and Institute for Theoretical Physics, Universit\"at zu K\"oln, D-50937
K\"oln, Germany. \texttt{as@thp.uni-koeln.de}}

\maketitle

\begin{abstract}
  We introduce an enhanced model based on the generalized centrifugal
  force model. Furthermore, the desired direction of pedestrians is
  investigated. A new approach leaning on the well-known concept of
  static and dynamic floor-fields in cellular automata is
  presented. Numerical results of the model are presented and compared
  with empirical data.
\end{abstract}

Force-based models try to describe the dynamics of pedestrians as
reaction to forces acting on each single pedestrian. Basically
two kinds of forces can be distinguished:
\begin{itemize}
\item {\em driving forces} designed to drive pedestrians to a desired
  direction with a desired speed.
\item  {\em repulsive forces} which are responsible for preserving the
  volume exclusion of pedestrians.
\end{itemize} 

Since the introduction of force-based models \cite{Helbing1991} many
works were dedicated to investigations of the repulsive forces and
finding new and better forms
\cite{Helbing1995,Lakoba2005,Yu2005,Johansson2007,Chraibi2010a}. These
efforts for improving the form of the repulsive force is understandable,
since the interactions between pedestrians dominate the dynamics,
especially at high densities. Surprisingly, not much work has been
done on the influence of the specific form of the driving force which
is expected to dominate the behavior at low or intermediate densities.

The standard form of the driving force is
\begin{equation}
\overrightarrow{F_{i}}^{\rm drv} = m_i \frac{\overrightarrow{v_{i}^0} -
 \overrightarrow{v_{i}}} {{\tau}},
\label{eq:fdrv}
\end{equation}
with a relaxation time ${\tau}$ and a  desired
velocity~$\overrightarrow{v_{i}^0}$. Although this expression 
is simple, it is not clear how to choose the desired direction
\begin{equation}
\overrightarrow{e_{i}^0}=\frac{\overrightarrow{v_{i}^0}}{\parallel
  \overrightarrow{v_{i}^0}\parallel}
  \label{eq:e0}
\end{equation}
in a given situation and only very few works were concerned with modeling the
desired direction (\ref{eq:e0}).  In \cite{Steffen2009} an Ansatz
with directing lines was introduced to steer pedestrians around
$90^\circ$\, and $180^\circ$ corners.  Gloor et al. \cite{Gloor2003}
used a path-oriented approach to model the desired direction of agents
on given hiking paths.
 
In \cite{Moussaid2011} Moussa\"id et al.\ have formulated the determination
of the desired direction in form of a minimization problem. 

It should be mentioned that the {\em directing problem} we discuss
here, i.e.\ the determination of the desired direction for each
pedestrian, is conceptually different from the classical {\em routing
problem}. In \cite{Hocker2010} an algorithm for generating
automatically a navigation graph in complex buildings in combination
with directing lines at corners was proposed.  Another algorithm for
way finding in buildings was proposed in \cite{KemlohWagoum2010}. Recently a
further development of the notion of the ``quickest path'' using a
non-iterative method to estimate the desired direction in the social
force model (SFM) was introduced \cite{Kretz2011}. The
main concern in this class of problems is how to define and connect
intermediate targets, in order to facilitate the evacuation of
pedestrians. By contrast, in the directing problem the existence of
such intermediate targets is in general assumed.

In this work we introduce enhancements of the generalized centrifugal
force model (GCFM) and investigate on their basis the modeling of the
desired direction (\ref{eq:e0}). For the sake of demonstration we
test our model in two different geometries: a bottleneck and a corner.


\section{The model}

In this section we give a brief overview of the GCFM and its
definition. Furthermore we introduce an effective modification of the
pedestrian-wall interactions that simplifies the definition of the
repulsive force.

\subsection{Pedestrian-pedestrian repulsive interactions}
\label{sec:ppi}

Introducing the vector connecting the positions of pedestrians $i$ and $j$,
\begin{equation}
   \overrightarrow{R_{ij}} = \overrightarrow{R_j}
   -\overrightarrow{R_i} ,\;\;\;\; \qquad
   \overrightarrow{e_{ij}} = \frac{\overrightarrow{R_{ij}}}{\parallel
\overrightarrow{R_{ij}}\parallel}\,,
\label{eq:R}
\end{equation} 
the repulsive force in the GCFM reads
\begin{equation}
   \overrightarrow{F_{ij}}^{\rm rep}=-m_i k_{ij}\frac{(\eta {v_i}^0 +
     v_{ij})^2}{d_{ij}} \overrightarrow{e_{ij}},
\label{eq:modCFMfrep3}
\end{equation}
with $m_i=1$ the mass of $i$ and the effective distance between pedestrian $i$ and $j$,
\begin{equation}
d_{ij} = \parallel\overrightarrow{R_{ij}}\parallel 
- r_i(v_i) -  r_j(v_j)  \,,
\label{eq:dist}
\end{equation}
and the polar radius $r_i$ of pedestrian~$i$.

The relative velocity $v_{ij}$ is defined such that slower pedestrians
are less affected by the presence of faster pedestrians in front of them:
\begin{eqnarray}
  v_{ij} &=&
  \frac{1}{2}[(\overrightarrow{v_i}-\overrightarrow{v_j})\cdot
  \overrightarrow{e_{ij}} +
  |(\overrightarrow{v_i}-\overrightarrow{v_j})
  \cdot\overrightarrow{e_{ij}}|] \nonumber \\ 
 &=&\begin{cases}
   (\overrightarrow{v_i}-\overrightarrow{v_j})\cdot
   \overrightarrow{e_{ij}} & \quad\mbox{if}\;\;
   (\overrightarrow{v_i}-\overrightarrow{v_j})\cdot
   \overrightarrow{e_{ij}} > 0 \\ 0 & \quad\mbox{otherwise.}
 \end{cases}
\end{eqnarray}
The parameter
\begin{eqnarray}
  \label{eq:kij}
  k_{ij} & = &
  \frac{1}{2}\frac{\overrightarrow{v_i}\cdot\overrightarrow{e_{ij}} +
    \mid \overrightarrow{v_i}\cdot\overrightarrow{e_{ij}} \mid} {v_i}
  \nonumber \\ 
  & = &
\begin{cases}
  (\overrightarrow{v_i}\cdot\overrightarrow{e_{ij}})/\parallel
  \overrightarrow{v_i}\parallel & \quad\mbox{if}\;\; 
   \overrightarrow{v_i}\cdot\overrightarrow{e_{ij}}>0\; \And\; 
    \parallel\overrightarrow{v_i}\parallel \ne 0\\ 
  0 & \quad\mbox{otherwise,}
 \end{cases}
\end{eqnarray}
reduces the effective range of the repulsive force to the angle of
vision.  Through the coefficient $k_{ij}$ the strength of the
repulsive force depends on the angle: it is maximal when pedestrian
$j$ is in the direction of motion of pedestrian $i$ and minimal when
the angle between $j$ and $i$ is bigger than $90^\circ$.

\subsection{Wall-pedestrian repulsive interactions}
\label{sec:wpi}
In the GCFM the interactions between pedestrians and walls are modeled
by a force similar to the pedestrian-pedestrian repulsive force. A wall is
represented by three point masses acting on pedestrians within a
certain range. From a computational point of view this analogy
exhibits an overhead since the repulsive force between a pedestrian
and a wall is calculated three times.

\begin{figure}
\begin{center}
\includegraphics[width=0.6\columnwidth]{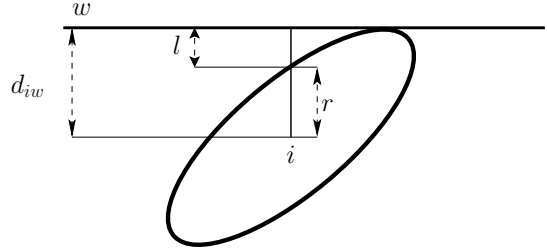}
\caption{Illustration of distances used in the definition of the wall-pedestrian repulsive force~(\ref{eq:rep-ped-wall}).}     
\label{fig:Frep-direction}
\end{center}
\end{figure}

We now make use of the ``distance of closest approach'' as defined in
\cite{Chraibi2010a} to formulate the repulsive force between a
pedestrian $i$ and a wall $w$ as
\begin{equation}
\overrightarrow{F_{iw}}^{\rm rep} = \eta^\prime \parallel \overrightarrow{{v_i}^0}\parallel k_{iw}  b_{iw},
\label{eq:rep-ped-wall}
\end{equation}
with 
\begin{eqnarray}
  \label{eq:bij}
  b_{iw} & = &
  H\left( 1-\frac{d_{iw}}{r+l}\right)\cdot\left( 1-\frac{d_{iw}}{r+l}
  \right),
\end{eqnarray}
where $l$ is the distance of closest approach between an ellipse and a
line, $r$ is the polar radius determined by the nearest point on the
line to the center of the ellipse $i$ (Fig.~\ref{fig:Frep-direction}). $H()$ is the Heaviside
step function, $k_{iw}$ is defined in Eq.(\ref{eq:kij}), $\parallel \overrightarrow{{v_i}^0}\parallel$ is the desired speed of $i$ and $\eta^\prime$ is a parameter to control the strength
of the force.

The repulsive force (\ref{eq:rep-ped-wall}) is a contact force
that is different from zero if the effective distance of the center of
the ellipse to the segment line is non-positive. For the simulations
in this paper we set the strength of the repulsive forces as 
$\eta=0.2$ and $\eta^\prime=5$.


\section{Influence of the desired direction}
\label{Chap3::influence}

\begin{figure}[H]
\begin{center}
\includegraphics[width=0.60\columnwidth]{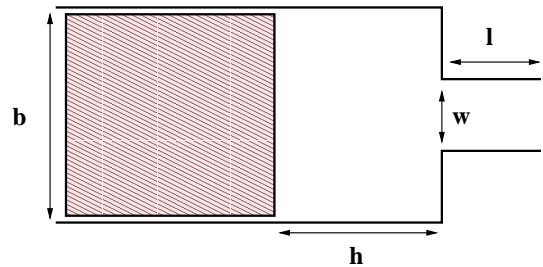}
\caption{Scenario set-up: Pedestrians move from a holding area 
  (shaded region) through the bottleneck ($l=2.8\,\rm{m}$,
  $h=4.5\,\rm{m}$, $b=4\,\rm{m}$ and $w$ variable).}
\label{fig:setup-bot}
\end{center}
\end{figure}

In this section we study the effects of the desired direction on the
dynamics of a system by measuring the outflow from a bottleneck with
different widths. See Fig.~\ref{fig:setup-bot} for the simulation set-up. 
Four different methods for setting the direction of
the desired velocity are introduced and discussed. Finally, simulation results will
be compared.
\subsection{Strategy 1: Directing towards the middle of the exit}

\begin{figure}[H]
 \begin{center}
  \includegraphics[width=0.8\columnwidth]{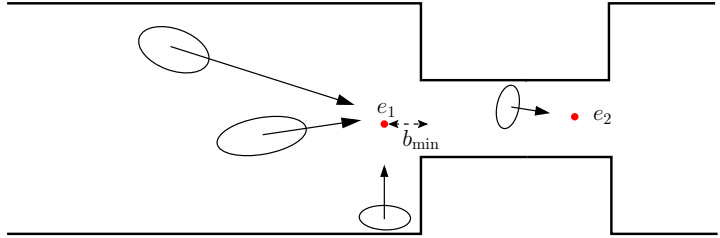}
  \caption{Strategy 1: All pedestrians are directed exactly 
    towards the reference points $e_1$\, and $e_2$.}
  \label{fig:str1}
 \end{center}
\end{figure}

The first strategy is probably the most obvious one. Herein, the 
desired direction $\overrightarrow{e_{i}^0}$  for pedestrian $i$ 
is permanently directed towards a reference point that exactly
lies on the middle of the exit. In some situations it happens that
pedestrians can not get to the chosen reference point without
colliding with walls. To avoid this and to make sure that all
pedestrians can ``see'' the middle of the exit the reference point
$e_1$ is shifted by half the minimal shoulder length $b_{\rm
  min}=0.2\,\rm{m}$ (Fig.~\ref{fig:str1}). Pedestrians that pass to the right of the reference point $e_1$ head towards $e_2$. 

\subsection{Strategy 2: Enhanced directing towards the middle of the exit}

This is a modification of strategy 1. Pedestrian are still directed to
the shifted reference point $e_1$. However, from a certain position
pedestrians can see through the bottleneck the second reference point
$e_2$. In this case $e_1$ is ignored and the desired direction is set
to be parallel to the line $\overrightarrow{e_1e_2}$. Since, pedestrians that are inside the bottleneck can always see $e_2$ the desired direction is kept parallel to $\overrightarrow{e_1e_2}$.

\begin{figure}[H]
 \begin{center}
  \includegraphics[width=0.8\columnwidth]{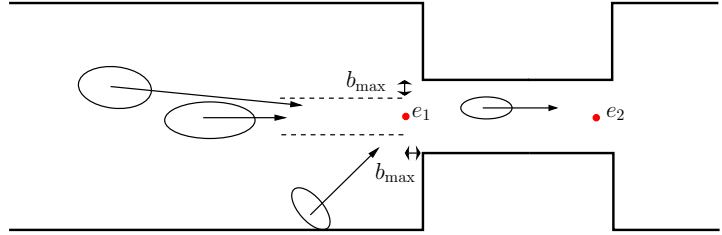}
  \caption{Strategy 2:  Depending on their position pedestrians adapt
    their direction. In the range where the exit of the bottleneck is
visible (marked by dashed lines) the direction is longitudinal.
Outside this area they are directed towards the middle of the bottleneck.
}
  \label{fig:str2}
 \end{center}
\end{figure}

Here again the reference points and the delimiting range of the
bottleneck is shifted in $x$- and $y$-direction by $b_{\rm max}$
(Fig.~\ref{fig:str2}).

\subsection{Strategy 3: Directing towards the nearest point on the exit}

\begin{figure}[H]
 \begin{center}
  \includegraphics[width=0.8\columnwidth]{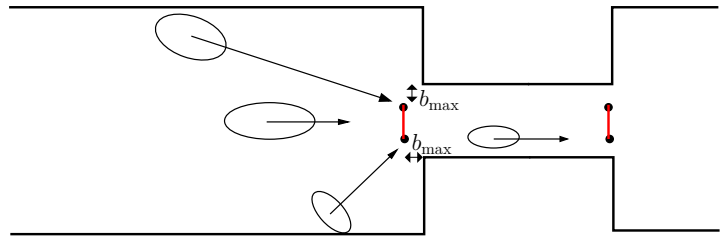}
  \caption{Strategy 3:  Directing towards the nearest point 
    on the exit. Moln\'ar published in \cite{Molnar1995} a very
    similar strategy. The only difference is the placement of the
    line, which is away from the corner by $b_{\rm max}$.  }
  \label{fig:str3}
 \end{center}
\end{figure}

Another possibility to choose the desired direction
$\overrightarrow{e_{i}^0}$ is to define a line $l$ in front of the
exit and take at each time the nearest point from the pedestrian $i$
to $l$ (Fig.~\ref{fig:str3}). In comparison with strategy 2,
pedestrians that are not in the range where the point $e_2$ is not
visible choose one of the end points of the line $l$. In strategy 2
this would be the middle of $l$.

\subsection{Strategy 4: Guiding line segments}
\label{subsec:str4}
\begin{figure}[H]
 \begin{center}
  \includegraphics[width=0.8\columnwidth]{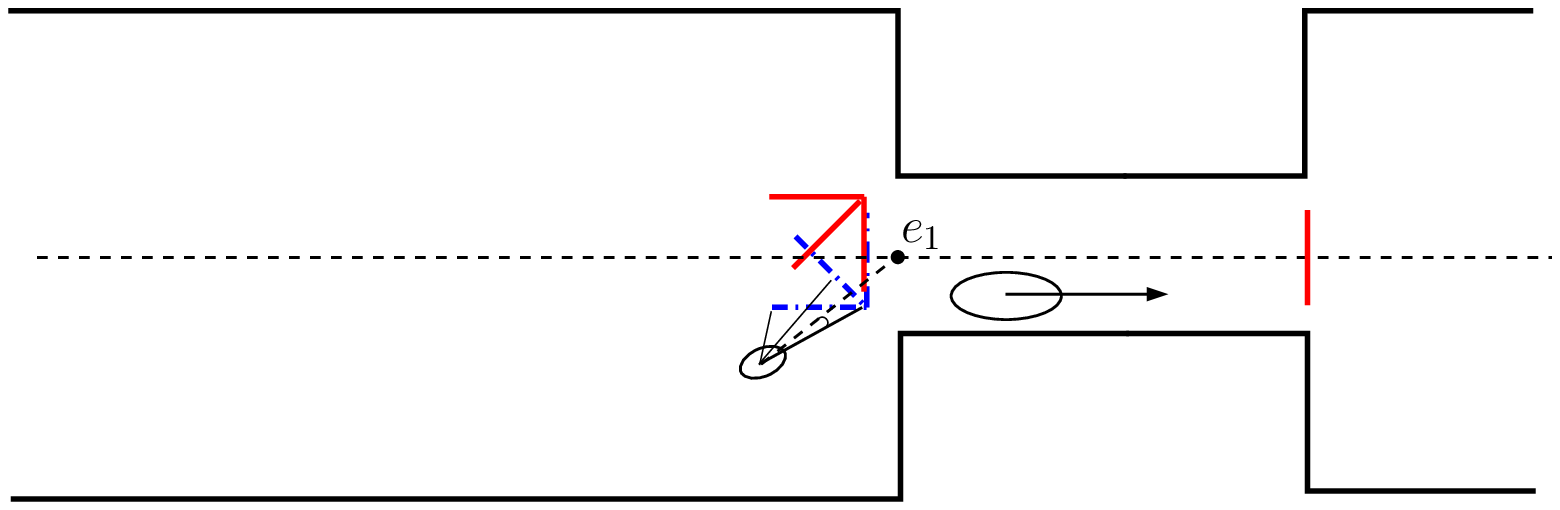}
  \caption{Strategy 4: Guiding line segments in front of the 
    bottleneck. For each corner a set of three line segments is
    generated. The length of all directing lines is equal to 3.5 m.}
  \label{fig:str4}
 \end{center}
\end{figure}

Without loss of generality we introduce the main idea of strategy 4
with help of the previous bottleneck.  Unlike the previous strategies
this strategy is applicable to all geometries with corners even if the
exit point is not visible. We recall that in strategy 3 a line in front of
the bottleneck was defined. The nearest point from each pedestrian to
this line was set to define the desired direction. As a generalization
we make use in strategy 4 of three different lines to ``smoothen'' merging
in front of the bottleneck (Fig.~\ref{fig:str4}). 
The blue line set (down the dashed line segment) is considered by
pedestrians in the lower half and the red
line set by pedestrians in the upper half of the
bottleneck. For a pedestrian $i$ at position $p_i$ we define the angle
\begin{equation}
\theta_i = \arccos \left(\frac{\overrightarrow{p_ie_1}\cdot 
\overrightarrow{p_il_{ij}}}{\parallel \overrightarrow{p_ie_1}\parallel
\cdot\parallel\overrightarrow{p_il_{ij}}\parallel}\right),
\end{equation}
with $l_{ij}$ the nearest  point of the line $j$ to the pedestrian $i$.

The next direction is then chosen as
\begin{equation}
\overrightarrow{e_{i}^0}=\frac{\overrightarrow{p_il_{ij}}}{\parallel
  \overrightarrow{p_il_{ij}}\parallel}
\end{equation}
with $j$ such that $\theta_j=\min\{\theta_1,\theta_2,\theta_3\}$. As in
strategy 3 the direction lines are shifted in $x$- and $y$-direction 
by $b_{\rm min}$.

\subsection{Numerical results}

In the previous section we have proposed different methods (called
strategies) for choosing the desired direction
$\overrightarrow{e_{i}^0}$.  To compare these strategies we have 
performed simulations for a bottleneck using the same set of 
parameters for the GCFM. For each strategy only the width of the
bottleneck was varied from $1$~m to $2.4$~m.

On the basis of a quantitative analysis the importance
of the choice of strategy for the observed behavior can be estimated.
In the following, for each strategy  we measure the flow through
bottlenecks of varying width $w$. The flow is measured directly at the
entrance of the bottleneck according to
\begin{equation}
        J = \frac{N_{\Delta t}-1}{\Delta t},
\end{equation}
with $N_{\Delta t}=60$ pedestrians and $ \Delta t$ the time necessary
that all pedestrians pass the measurement line.

\begin{figure}[H]
\begin{center}
\includegraphics[width=0.95\columnwidth]{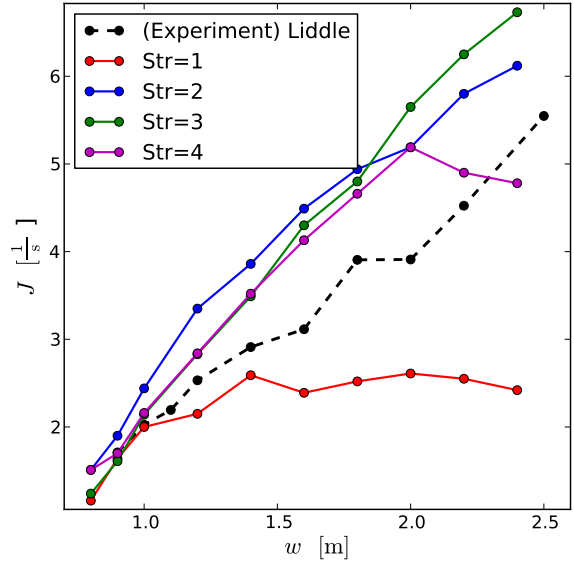}
\caption{Flow through a bottleneck with different widths. Simulation 
  results with different strategies for the desired direction of
  pedestrians in comparison with empirical data from
  \cite{Liddle2009}. The experiments were conducted with 180 persons.}
\label{fig:flow-bot}
\end{center}
\end{figure}

In Fig.~\ref{fig:flow-bot} the resulting flow for all four strategies
is presented.
The flow for strategy 1 saturates independently of the width. This was
expected since pedestrians do not use the whole width of the
bottleneck and keep indeed oriented to the middle. The picture changes
for strategies 2 -- 4, where the effective width of the bottleneck
is clearly larger.  Strategy 2 shows a better usage of the middle
widths ($\leq 1.8 {\rm m}$).  Here, the slight blocking near the
corners, that emerges from strategy 3 is particularly disadvantageous.
Strategy 4 produces higher flows for widths up to $2\, {\rm m}$. Up
$2\, {\rm m}$ the flow stagnates.
The main observations are:
\begin{itemize}
\item The choice of the strategy for the desired direction influences
  considerably the outcome of the simulation.
\item An inconsiderate choice of strategy, in that
  case strategy 1, can lead to large variations from experimental results.
\item In contrast to strategy 1, strategies 2 -- 4 show better usage of
  the bottleneck width and lead to higher flow values.
\end{itemize}


\section{An application: Motion around a corner}

Basically, force-based models are functional only in areas, where the exit is constantly visible by all pedestrians.
Obviously, this can not always be guaranteed which is a problem since
a proper initialization of the desired direction
$\overrightarrow{e_{i}^0}$ for each pedestrian is not possible. In
order to overcome this problem one has to introduce ``virtual'' exits.
This was showcased previously with strategy 4.

In this section, we introduce enhancements of strategy 4 and study
their impact on the movement time, i.e. the time until all
pedestrians have left the simulation set up. For simplicity we
consider the movement of $N=100$ pedestrians in a $90^\circ$-corner-like corridor.

The basis of our enhancements is the following observation: Given a
guiding line $l$, the desired direction of a pedestrian $i$ is
determined in dependence of its position and the nearest point to $l$.
This choice neglects two important factors:
\begin{enumerate}
\item The perception of space: Individuals try to
  minimize, when possible, their path to the exit. In our example,
  pedestrians would take a point near the corner as goal and not the
  nearest point on the guiding line. Depending on the starting
  position of pedestrians, this can be far away from the corner and
  much longer than the shortest path to the exit.
\item The dynamical and collective influence of pedestrians: In the
  presence of other pedestrians and depending on the magnitude of the
  local density, the nature of the ``quickest path'' \cite{Kretz2009}
  changes dynamically and differs in most cases from the ``shortest
  path'' to the exit.
\end{enumerate} 

We therefore adopt a concept similar to ideas introduced in
\cite{Burstedde2001} which are well established and widely used in
cellular automata models \cite{Nishinari2004,Kretz2006c}.
 
At a time step $t$ a pedestrian $i$ heads towards a point on the line 
which minimizes the distance to the inner point of the corner $l^i(t)$. This is a natural
territorial effect which leads to the shortest path to the exit. If
all pedestrians try to take the shortest path, large jams will be
observed right at the inner point of the corner. If, however, the
collective influence of pedestrians dominates the choice of the
desired direction, pedestrians will choose their desired direction to
be orthogonal to the guiding lines and thus make better use of the
whole directing line.

\begin{figure}[H]
\begin{center}
\includegraphics[width=0.55\columnwidth]{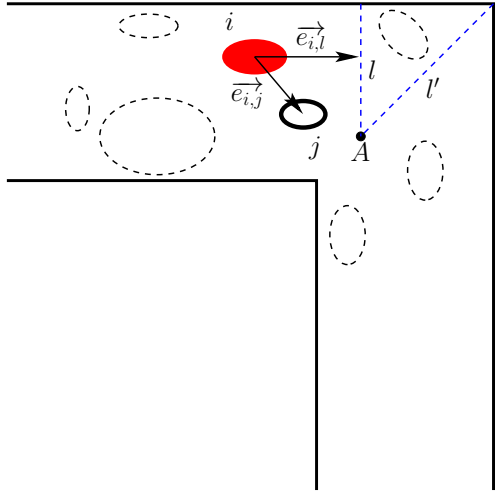}
\caption{How to get around the corner? Pedestrian $i$ that is heading 
  toward the first guiding line, considers the positions of its
  neighboring pedestrians as well as its initial position to decide
  whether or not to head closer to the edge of the corner.}
\label{fig:relevantpeds}
\end{center}
\end{figure}

For this reason we include a dynamical factor that depends mainly on
previous decisions taken by other pedestrians:
\begin{equation}
 p^i(t) =  \exp\left(-k_d\cdot {\rm occ}^i_{\rm rel}(t)\right),
\label{eq:dynFF}
\end{equation} 
where 
\begin{equation}
{\rm occ}^i_{\rm rel}(t)=\frac{n^i}{n^i_{\rm max}}
\label{eq:occrel}
\end{equation} 
is a measure of the occupation of the line.  $n^i$ is the cardinality
of the set $$A_l=\big\{ l^j\; |\; j\in B_l\; \&\; l^j<l^i \big\}$$ and $n^i_{\rm max}$
is the cardinality of the set
\begin{equation}
  B_l=\big\{ j \in [1,N] \;|\; i \neq j \And
  \overrightarrow{e_{i,l}}\cdot \overrightarrow{e_{i,j}} \geq 0
  \big\}.
\end{equation} 
$B_l$ is the set of all relevant neighbors of $i$, that influences its desired direction by means of a contribution to  $\rm{occ}^i_{\rm rel}(t)$ (\ref{eq:occrel}). For the scenario depicted in Fig.~\ref{fig:relevantpeds} the set $B_l$ for $i$ (red ellipse) contains  only one pedestrian $j$ (bold ellipse).

Large values of ${\rm occ}^i_{\rm rel}(t)$ imply small values of
$p^i(t)$. As a consequence pedestrians prefer not to change their
desired direction closer to the edge of the corner.

Finally, the update rule of the distance $l^i(t)$ is given by:
\begin{equation}
l^i(t+\Delta t) =  l^i(t)\cdot \Big(1-p^i(t) \Big).
\label{eq:lneu}
\end{equation}
$p^i(t) \in [0,1]$ gives the rate of change from the initial ``guess''
of pedestrian $i$. For $p^i(t)=0$ the desired direction of $i$ stays
orthogonal to the guiding line, while $p^i(t)=1$ displays the case
where $i$'s desired direction is directed to the edge of the corner $A$. In the next section  we study the influence of the parameter $k_d$ on the dynamics of pedestrians. For the second and third line we set $k_d=0$ and vary it only for the first line.
\section{Analysis of the sensitivity parameter}

To understand the impact of the collective influence of pedestrians on
the chosen target point for each pedestrian $i$, we study the time
evolution of the relative length for different values of $k_d$.
The relative length is defined as 
\begin{equation}
l^i_{\rm rel}(t)=\frac{l^i(t)}{l_{\rm max}}
\label{eq:lrel}
\end{equation} 
where $l_{\rm max}$ is the length of the guiding line.

\begin{figure}
  \centering
  \subfloat[$k_d=0$]{\label{fig:ks-kd0-0}\includegraphics[width=0.53\textwidth]{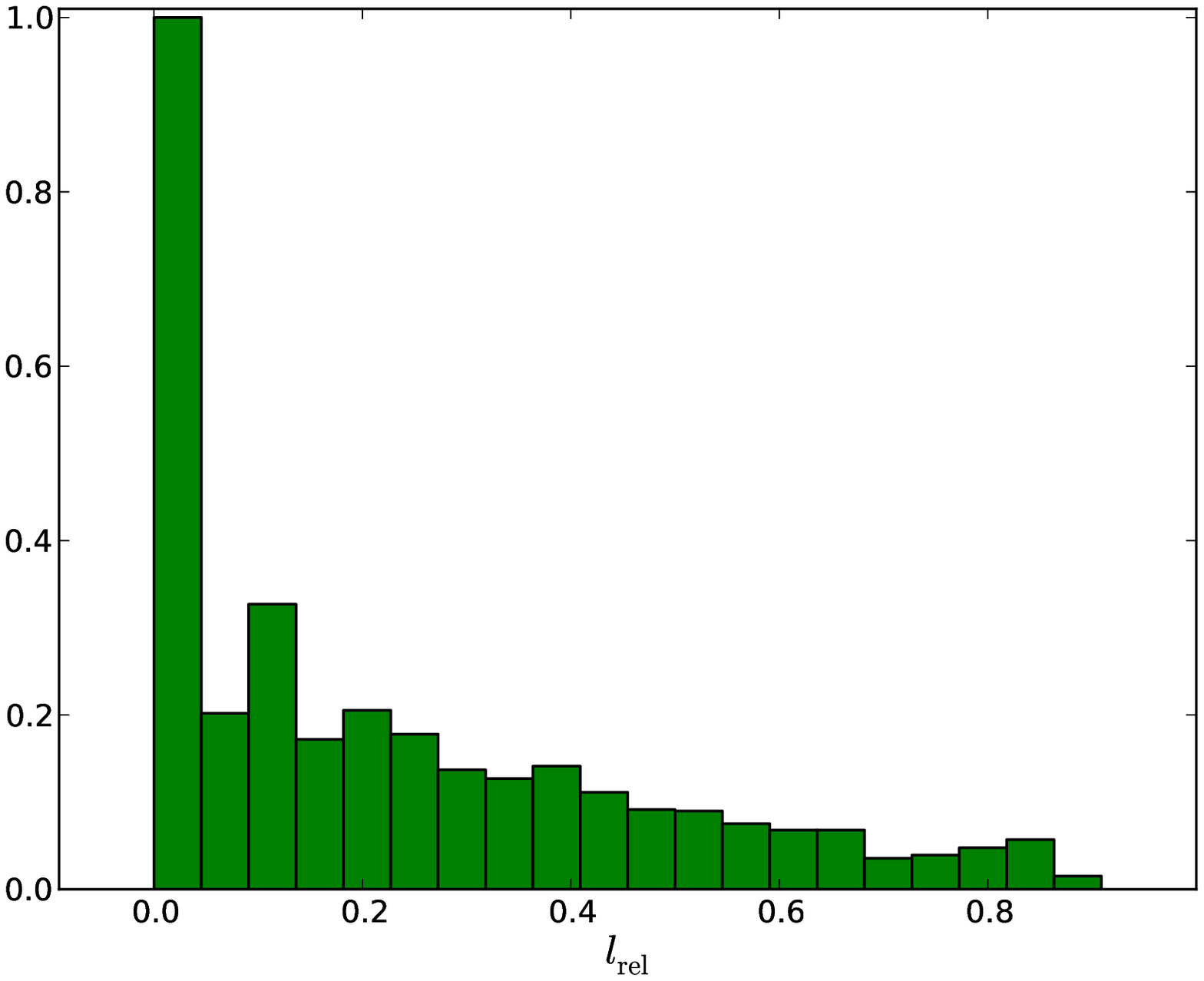}}                
  \subfloat[$k_d=10$]{\label{fig:ks-kd0-1}\includegraphics[width=0.53\textwidth]{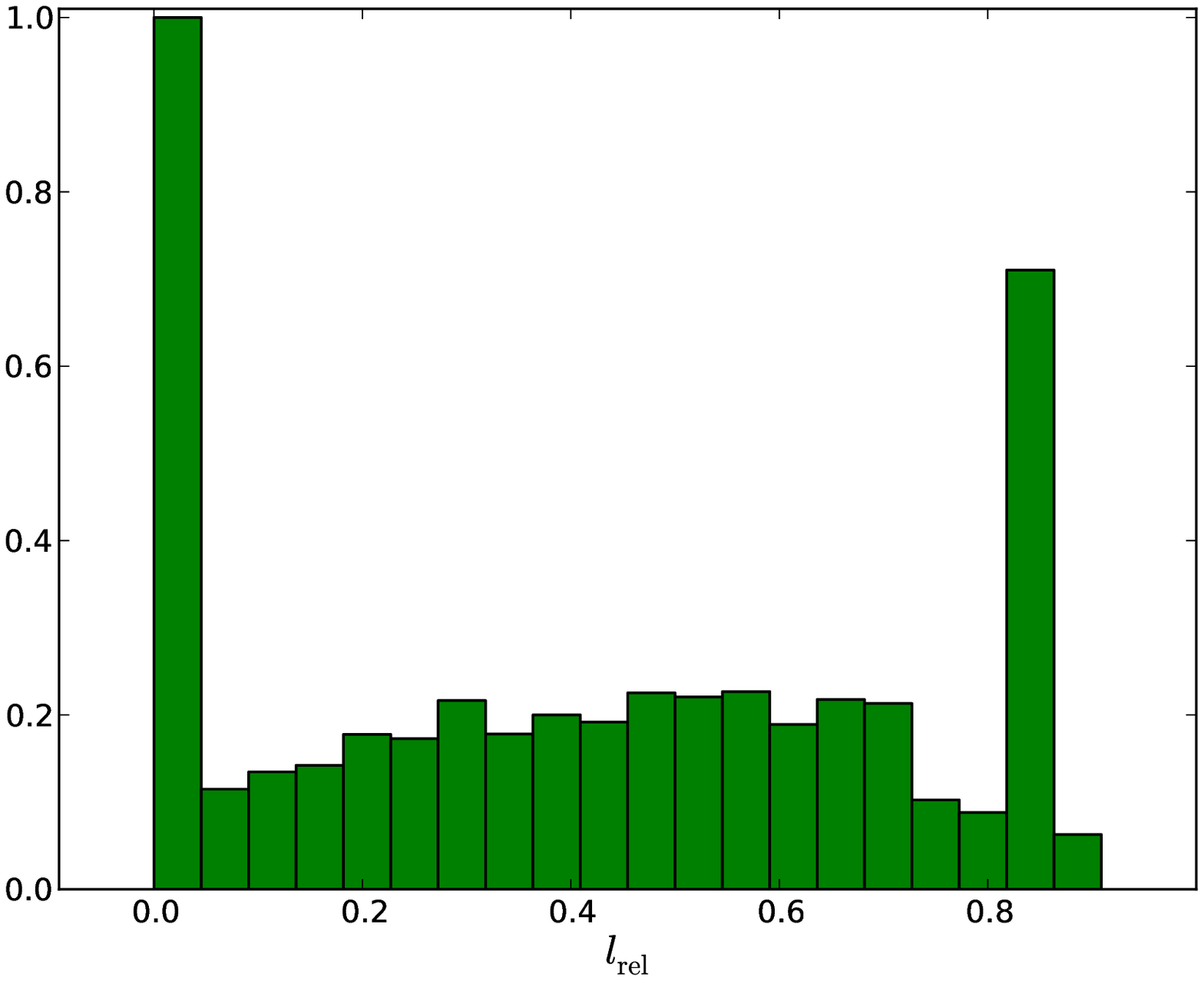}}
  \caption{Movement of $N=100$ pedestrians around a corner with different values of $k_d$. The width of the corridor is $w=3$~m.}
  \label{fig:kd0-10}
\end{figure}

Fig.~\ref{fig:ks-kd0-0} shows the probability distribution of the
relative length for $k_d=0$. Pedestrians are mainly heading towards
$A$ and the full length of the directing line is rarely used.

The situation changes considerably for $k_d = 10$.
Fig.~\ref{fig:ks-kd0-1} shows that the distribution of the length is
more balanced which indicates that pedestrians make better use
of the directing lines.

\begin{figure}[H]
\begin{center}
\includegraphics[width=0.64\columnwidth]{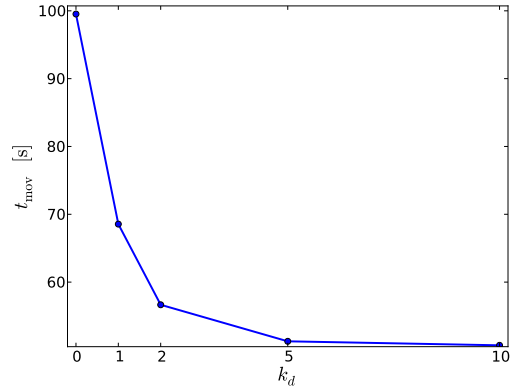}
\caption{Movement time for a simulation with $N=100$ pedestrians around 
  a corner for different values of $k_d$. }
\label{fig:evactime}
\end{center}
\end{figure}

\begin{figure}[H]
\begin{center}
 \subfloat{}\includegraphics[width=0.66\columnwidth]{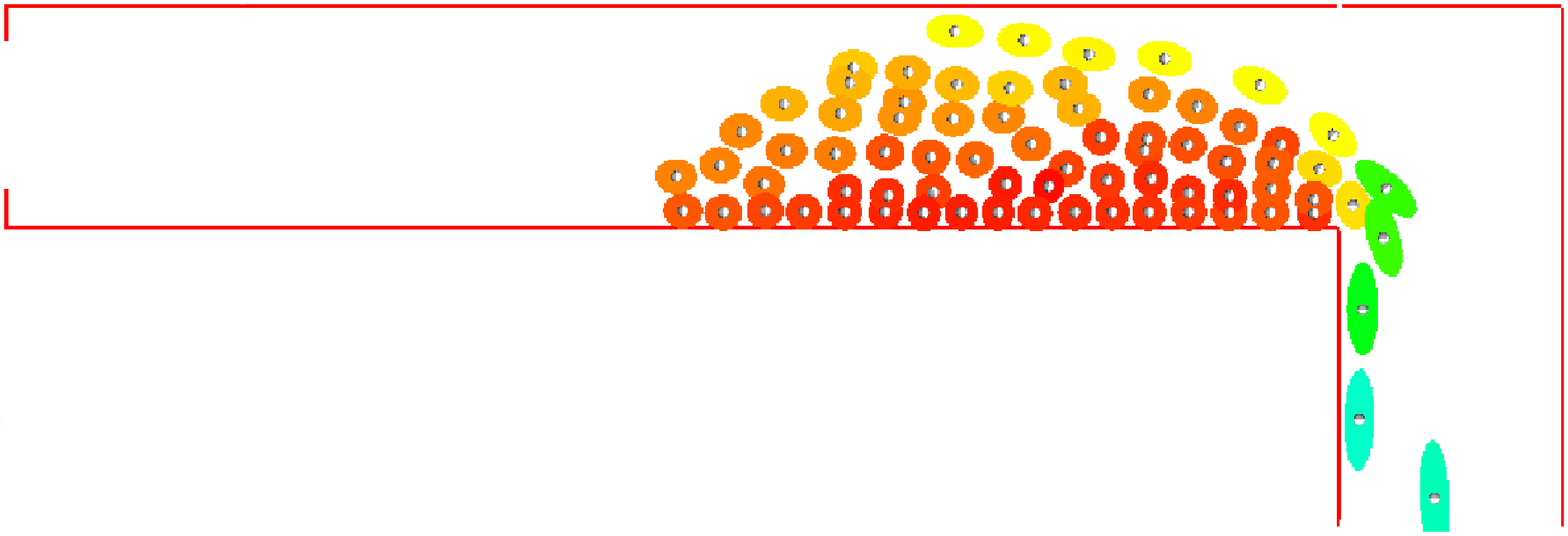}
\subfloat{}\includegraphics[width=0.66\columnwidth]{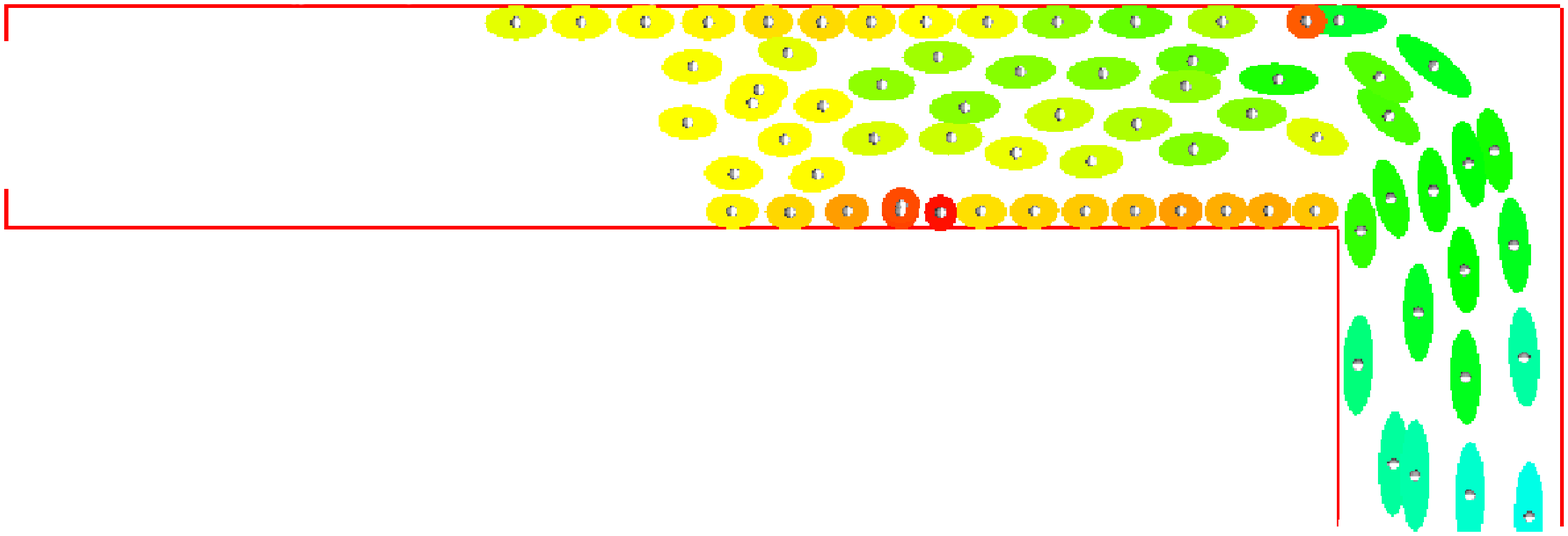}
\caption{Screen shot of a simulation with 100 pedestrians, $k_d=0$
(top) and  $k_d=10$ (bottom).
  }
\label{fig:jamNojam}
\end{center}
\end{figure}

To showcase the impact of collective influence of pedestrians on the
desired direction, we show in Fig.~\ref{fig:evactime} the variation of
the movement time in dependence of $k_d$.
A qualitative comparison shown in Fig.~\ref{fig:jamNojam} confirms the
above-mentioned quantitative analysis.

For $k_d=0$ a jam forms immediately before the corner as indicated by
the large number of slowly moving pedestrians (red ellipses). This
results from a strong competition between the pedestrian to pass close
to the edge $A$ of the corner. In contrast, for $k_d=10$ pedestrians move
quicker since they make optimal use of the guiding line.


\section{Summary}

We have developed a strategy to determine the desired direction
$\overrightarrow{e_i^0}$ for each pedestrian $i$. This method is
rather general and can be used in each geometry characterized by the
existence of corners, e.g. bottlenecks (2 corners), T-Junction (2
corners).  In analogy to CA models we introduced and tested a factor
to model the static and dynamic interactions of pedestrians with the
geometry.

Our work was based on an enhanced version of the GCFM
\cite{Chraibi2010a}. The enhancements use a considerable
simplification of the repulsive forces acting on pedestrians from
walls. Furthermore, we addressed an important issue in force-based
models, namely the choice of the desired direction of
pedestrians. Several strategies were implemented and compared with
empirical data. This comparative investigation showed that the outcome
of a simulation depends strongly on the chosen direction of the
desired direction of pedestrians.  Finally, we introduced a new
mechanism to direct pedestrians in $90^\circ$-corners by means of
directing lines. The main concept of this strategy base on the
well-known concept of dynamical floor-field. For further works, the
parameter $k_d$ that expresses the tendency of pedestrians to take the
shortest path (or not) should be varied individually as the geometrical
and dynamical conception of pedestrians differ. 


\section*{Acknowledgments} 

This work is within the framework of two projects.  The authors are
grateful to the Deutsche Forschungsgemeinschaft (DFG) for funding the
project under Grant-No.~SE 1789/1-1 as well as the Federal Ministry
of Education and Research (BMBF) for funding the project under
Grant-No.~13N9952 and 13N9960.



\printindex
\end{document}